\documentclass[iop,apj]{emulateapj}
\usepackage{float}
\usepackage{amsmath}
\usepackage{rotating}
\usepackage{amssymb}
\usepackage{hyperref}

\def\gtrsim{\mathrel{\hbox{\rlap{\hbox{\lower4pt\hbox{$\sim$}}}\hbox{$>$}}}}
\def\lesssim{\mathrel{\hbox{\rlap{\hbox{\lower4pt\hbox{$\sim$}}}\hbox{$<$}}}}
\def\gtrsim{\mathrel{\hbox{\rlap{\hbox{\lower4pt\hbox{$\sim$}}}\hbox{$>$}}}}
\def\farcs{\hbox{$.\!\!^{\prime\prime}$}}
\def\farcm{\hbox{$.\!\!^{\prime}$}}

\begin{document}

\def\chan{{\sl CXO\ }}

\title{The Variable Pulsar Wind Nebula of PSR J1809--1917}

\author{Noel Klingler$^{1,2}$, Oleg Kargaltsev$^{1,2}$, George G. Pavlov$^3$, Bettina Posselt$^3$}
\affil{
$^1$The George Washington University, Department of Physics, 725 21st St NW, Washington, DC 20052, USA \\ 
$^2$Astronomy, Physics and Statistics Institute of Sciences (APSIS), The George Washington University, Washington, DC 20052, USA \\
$^3$Pennsylvania State University, Department of Astronomy \& Astrophysics, 525 Davey Laboratory, University Park, PA 16802, USA
}

\begin{abstract}
PSR J1809--1917 is a young ($\tau=51$ kyr) energetic ($\dot{E}=1.8\times10^{36}$ erg s$^{-1}$) radio pulsar powering a pulsar wind nebula (PWN). 
We report on the results of three {\sl Chandra X-ray Observatory} observations which show that the PWN consists of a small ($\sim 20''$) bright compact nebula (CN) and faint extended emission seen up to $2'$ from the pulsar.  
The CN is elongated in the northeast-southwest direction and exhibits morphological and flux variability on a timescale of a few months. 
We also find evidence of small arcsecond-scale jets extending from the pulsar along the same direction, and exhibiting a hard power-law (PL) spectrum with photon index $\Gamma_{\rm jet}=1.2\pm0.1$. 
The more extended emission and CN share the same symmetry axis, which is also aligned with the direction toward the TeV $\gamma$-ray source HESS J1809--193, supporting their association. 
The spectrum of the extended nebula (EN) fits an absorbed PL with
about the same slope as that of the CN, $\Gamma_{\rm CN}\approx\Gamma_{\rm EN}=1.55\pm0.09$; no spectral changes across the EN's 2 pc extent are seen.
The total PWN 0.5--8 keV luminosity is $L_{\rm PWN}\approx 9\times10^{32}$ erg s$^{-1}$, about half of which is due to the EN.
\end{abstract}


\section{INTRODUCTION}
Pulsars are among the most powerful particle accelerators, capable of producing particles up to PeV energies. 
As a neutron star rotates, its rotational energy is imparted to an ultra-relativistic magnetized particle wind in the magnetosphere. 
Outside of the magnetosphere, the wind particles pass through the termination shock (TS), beyond which the gyration of particles in the magnetic field produces synchrotron radiation from radio to MeV $\gamma$-rays, and upscatters photons to produce inverse-Compton (IC) radiation from GeV to TeV $\gamma$-rays, which we can see as a pulsar wind nebula (PWN).

When pulsars are formed, they often acquire a substantial kick velocity from their parent supernova. 
Average 3D pulsar velocities have been found to be $v\sim400$ km s$^{-1}$ (Hobbs et al.\ 2005)  although the velocity distribution might be bimodal (see Verbunt et al.\ 2017). 
This means that most pulsars only remain in their parent supernova remnant (SNR) for a few tens of kiloyears. 
While inside SNRs, pulsar speeds are lower than the sound speed of the hot SNR interior, and the anisotropy of the pulsar wind manifests itself in the formation of structures such as jets and tori (Kargaltsev \& Pavlov 2008). 
Compared to the SNR interior, the ISM has a much lower sound speed $c_s \sim 3$-$30$ km s$^{-1}$ (depending on the ISM phase), so pulsars that have left their host SNR are usually moving supersonically. 
The ram pressure exerted by the ISM modifies the appearance and emission properties of the PWN by forming a bow shock with the apex ahead of the moving pulsar.
The bow shock confines the wind to the direction opposite that of the pulsar motion, and leads to the formation of an extended tail behind the pulsar (see Kargaltsev et al.\ 2017 and Reynolds et al.\ 2017 for recent reviews).

For several PWNe, repeated observations with the {\sl Chandra X-ray Observatory} ({\sl CXO}) revealed variability in brightness and structure. 
For instance, the torus of the Crab PWN consists of a moving pattern of multiple wisps (seen in radio, optical, and X-rays) which originate at the TS and then move outward with a projected velocity of $\sim0.3c$--$0.5c$ (Hester et al.\ 2002; Bietenholz et al.\ 2004). 
The morphological changes in the southern jet of the Crab PWN can be described as kinks propagating along the jet (Weisskopf 2011). 
Similar changes are seen in the northwestern (outer) jet of the Vela PWN, which also features blobs of plasma moving outward with apparent speeds 0.3--0.6 $c$ (Pavlov et al.\ 2003; Durant et al.\ 2013). 
Although the bright inner part of the Vela PWN does not appear to have a wispy structure similar to that of the Crab's torus, the surface brightnesses and shapes of the arcs and inner jets show modest variations on month timescales (Pavlov et al.\ 2001; Levenfish et al.\ 2013; Rangelov et al.\ 2015). 
The Kes 75 PWN exhibited more dramatic variability, possibly associated with a magnetar-like flares, with the flux of the inner jet changing by a factor of 2, and the brightness peak of the northern clump broadening and shifting outward by $\approx 2''$ on a timescale of 6 years (see Ng et al.\ 2008 and Livingstone et al.\ 2011). 
The additional more recent observations have shown that the northern jet's structure and surface brightness have evolved substantially over the 18-year span of observations, while the structure and brightness of the southern jet has remained roughly constant (Reynolds et al.\ 2018). 
Another PWN with jets showing clear variability in X-rays is that associated with the young PSRs B1509--58 in MSH15--52.
For example, the southeastern jet of PSR B1509--58, which extends 12 pc from the pulsar, varies in brightness by up to $30\%$ while exhibiting morphological variability on timescales of months-years, and shows hints of helicity similar to the Vela and Crab jets (see DeLaney et al.\ 2006).

Variability has been detected not only in young PWNe inside SNRs but also in older ones, that have left their parent SNRs.
Klingler et al.\ (2016b) and Posselt et al.\ (2017) presented evidence of variable emission knots in two PWNe associated with supersonically-moving pulsars, J1509--5850 and Geminga, respectively.

The high-energy particles that emit synchrotron radiation in X-rays cool on relatively short timescales ($\lesssim1$ kyr for typical PWN magnetic fields $\gtrsim 10$ $\mu$G), but the lower-energy particles, which emit inverse-Compton (IC) radiation in the GeV/TeV range, cool on much longer timescales ($\sim10-100$ kyr, for typical ISM photon densities). 
These longer-living IC-emitting particles can diffuse and/or advect far beyond the extent of the synchrotron-emitting regions, by factors of 10--100, forming extended TeV sources (sometimes called relic plerions). 
The sizes of TeV PWNe generally increase with pulsar age, while the opposite trend usually applies to the sizes of X-ray nebulae (see Kargaltsev et al.\ 2013). 
Because they are produced by older particles which have diffused or advected far from the pulsar, these TeV sources are sometimes positionally offset from their X-ray sources by a few to tens of parsecs, which can complicate establishing their associations to X-ray PWNe.

PSR J1809--1917 (hereafter, J1809) is a young and energetic radio pulsar with a characteristic age $\tau_{\rm sd}=P/(2\dot{P})=51$ kyr, spin-down power $\dot{E}=1.8\times10^{36}$ erg s$^{-1}$, and period $P=82.7$ ms (more parameters are provided in Table 1). 
J1809 is not associated with any known SNR, implying that it has either left its host SNR, or that its SNR is too dissipated and cold to be detected in X-rays or radio (see Section 4.2). 
Its dispersion measure (DM) of 197 pc cm$^{-3}$ implies a distance of 3.3 kpc when using the Yao, Manchester, \& Wang (2017) Galactic free electron density model\footnote{This distance is only slightly less than the 3.55 kpc estimate using  the Cordes and Lazio (2002) Galactic free electron density model.}. 
We adopt this distance everywhere below and scale all distant-dependent quantities to $d=3.3$ kpc. 
J1809 also lies $\sim7'$ ($\sim7$ pc) north-northeast of the TeV source HESS J1809--193, which is believed to be the PWN's TeV counterpart (HESS Collaboration 2017). 
Using a 20 ks {\sl Chandra} observation in 2004 (ObsID 3853), Kargaltsev \& Pavlov (2007; hereafter KP07) discovered a compact elongated X-ray PWN and fainter emission extending a few arcminutes from the pulsar, with a spectrum described by an absorbed power-law (PL) model with photon index $\Gamma=1.4\pm0.1$. 
We observed this target again in 2013 and 2014.
In this paper we report on the results of three \chan observations of PSR J1809--1917 and its PWN, which revealed variability in the compact nebula (CN) morphology on timescales of a few months, making the J1809 PWN a valuabe addition to the small sample of variable PWNe. 
In Section 2 we describe the observations and data reduction.  
In Section 3 we present the images and spectral fit results.  
In Section 4 we discuss our findings, and in Section 5 we summarize the results.

\begin{deluxetable}{lc}
\tablecolumns{9}
\tablecaption{Observed and Derived Pulsar Parameters \label{tbl-parameters}}
\tablewidth{0pt}
\tablehead{\colhead{Parameter} & \colhead{Value} }
\startdata
R.A. (J2000.0) & 18 09 43.147(7)  \\
Decl. (J2000.0) & --19 17 38.1(13)  \\
Epoch of position (MJD) & 51506  \\
Galactic longitude (deg) & 11.09  \\
Galactic latitude (deg) & --0.08  \\
Spin period, $P$ (ms) & 82.7  \\
Period derivative, $\dot{P}$ (10$^{-14}$) & 2.553 \\
Dispersion measure, DM (pc cm$^{-3}$) & 197.1  \\
Distance, $d$ (kpc) & 3.3  \\
Surface magnetic field, $B_s$ (10$^{12}$ G) & 1.5  \\
Spin-down power, $\dot{E}$ (10$^{36}$ erg s$^{-1}$) & 1.8  \\
Spin-down age, $\tau_{\rm sd} = P/(2\dot{P})$ (kyr) & 51.3 
\enddata
\tablenotetext{}{Parameters are from the ATNF Pulsar Catalog (Manchester et al.\ 2005).  The distance estimate is based on the dispersion measure listed and the Yao, Manchester, \& Wang (2017) Galactic free electron density model.}
\end{deluxetable}

\begin{deluxetable}{ccclc}
\tablecolumns{4}
\tablecaption{\chan observations used in our analysis
\label{tbl-obs}}
\tablewidth{0pt}
\tablehead{\colhead{ObsId} & \colhead{Instrument} & \colhead{Exposure,} & \colhead{Date} & \colhead{$\theta$, } \\ \colhead{} & \colhead{} & \colhead{ks} & \colhead{} & \colhead{arcmin}  }
\startdata
3853 & ACIS-S & 19.70 & 2004 Jul 21 & $0\farcm6$ \\
14820 & ACIS-I & 46.75 & 2013 Sep 29 & $0\farcm6$ \\
16489 & ACIS-I & 64.86 & 2014 May 25 & $1\farcm3$
\enddata
\tablenotetext{}{$\theta$ is the angular distance between the pulsar and the optical axis of the telescope.}
\end{deluxetable}

\section{OBSERVATIONS AND DATA REDUCTION}
We analyzed three \chan observations of J1809, with a total exposure of 131.31 ks.
The first observation (ObsID 3853; PI Sanwal) was conducted with the ACIS-S instrument while the latter two (ObsIDs 14820 and 16489; PI Pavlov) were conducted using ACIS-I. 
For all observations the ACIS detector was operated in VFAINT timed exposure mode, with a time resolution of 3.24 s.
The details of these observations are provided in Table 2.
The two other archival {\sl Chandra} observations of J1809, ObsIDs 6720 and 16475, were not included in our analysis because the pulsar and the compact nebula were far off-axis ($\sim$9$'$) in the former observation, while the latter observation (16475) was very short (2.4 ks).

We processed the data using the {\sl Chandra} Interactive Analysis of Observations (CIAO) software version 4.9 and the {\sl CXO} Calibration Data Base (CALDB) version 4.7.3.
All observations were reprocessed with {\tt chandra\_repro} to apply the latest calibrations.

To correct for possible systematic offsets in the \chan aspect solution, we performed astrometric corrections to the images.
We used the CIAO routine {\tt wavdetect} (Freeman et al.\ 2002) to identify the centroid positions of field X-ray point sources within $3\farcm$1 of the pulsar. 
We used ObsID 16489, the observation with the longest exposure, as the reference observation (to which the others will be aligned) since it provides the best centroiding accuracy. 
In ObsIDs 3853 and 14820, {\tt wavdetect} found 10 and 14 (respectively) field point sources which were also present in the reference observation (ObsID 16489). 
We then applied the {\tt wcs\_update} tool which minimizes the positional shifts of field point sources between the reference observation and each of the other two.
The R.A./Decl.\ correctional shifts of ObsIDs 3853 and 14820 (with respect to the reference observation, ObsID 16489) were $-0\farcs46$/$0\farcs44$ and $-0\farcs41$/$1\farcs15$, respectively.

After removing all field point sources from the astrometry-corrected images, we used the {\tt merge\_obs} script to combine the exposure-map-corrected images from each observation to study the faint extended emission. 
For all images and spectra we restricted the photon energies to the 0.5--8 keV range (unless otherwise noted).  
All uncertainties provided below are at the 1$\sigma$ confidence level.

\section{ANALYSIS AND RESULTS}

\subsection{PWN}

\subsubsection{Morphology}

\begin{figure*}
\epsscale{1.15}
\plotone{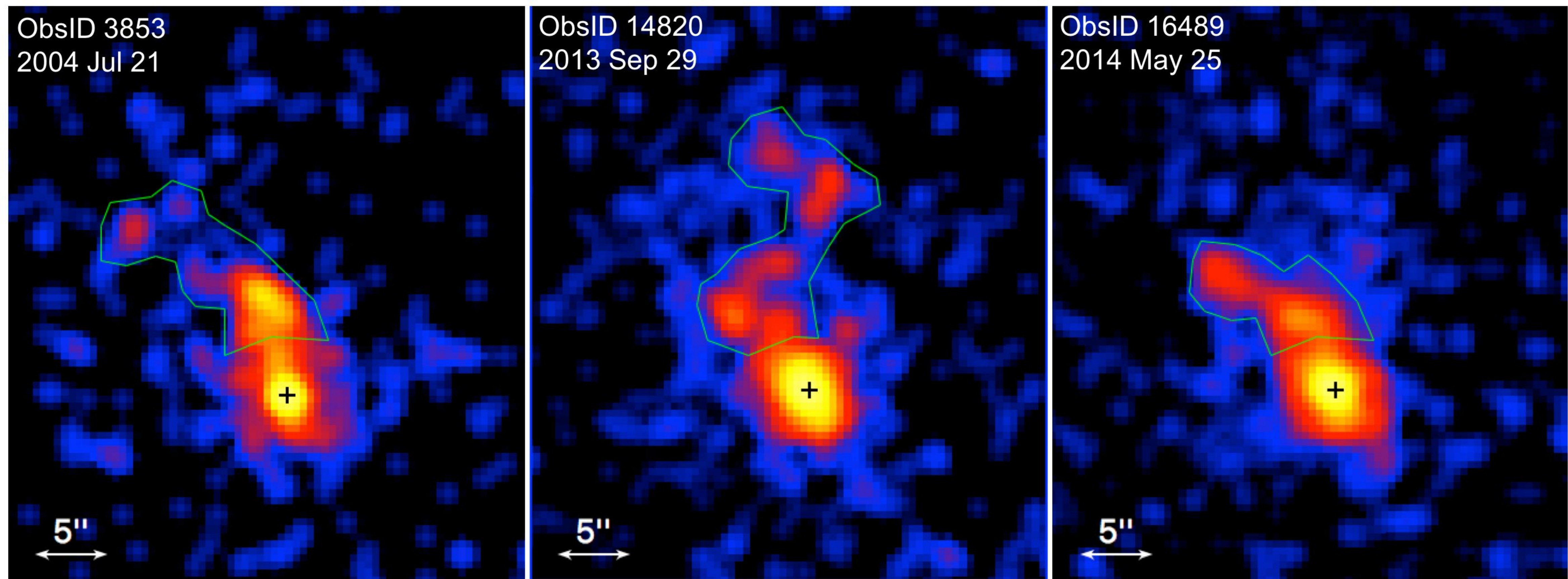}
\caption{Exposure-corrected {\sl Chandra} ACIS images of the J1809 PWN, smoothed with a 3-pixel ($r=1\farcs48$) Gaussian kernel, highlighting the  changes in the CN.  The green polygons are the  extraction regions  to look for spectral variability (see Section 3.2).  The pulsar position is shown by the cross.  North is up and East is to the left.}
\label{fig-variability}
\end{figure*}

Figure \ref{fig-variability} shows the high-resolution ACIS images of the compact nebula (CN) from the three {\sl CXO} observations. 
The CN is elongated in the northeast-southwest direction, approximately along a line with a position angle of about $30^\circ$ (counted east from north), with the northeast (NE) extension (up to $\approx$12$''$ from the pulsar; see the contours in Figure 1) appearing roughly three times longer than the southwest (SW) extension (seen only up to $\approx$4$''$).   
The NE extension of the CN is clearly variable on timescales less than 8 months (the interval of time between the two most recent observations). 
In ObsID 14820, the morphology of the brightest part of the NE extension resembles a double-kinked jet (cf.\ the Vela PWN jet images shown in Figure 4 from Pavlov et al.\ 2003). 
However, in the latest observation image, the farthest part of the ``jet''  is no longer visible while the part closest to the pulsar appears to have brightened.

\begin{figure*}
\epsscale{1.15}
\plotone{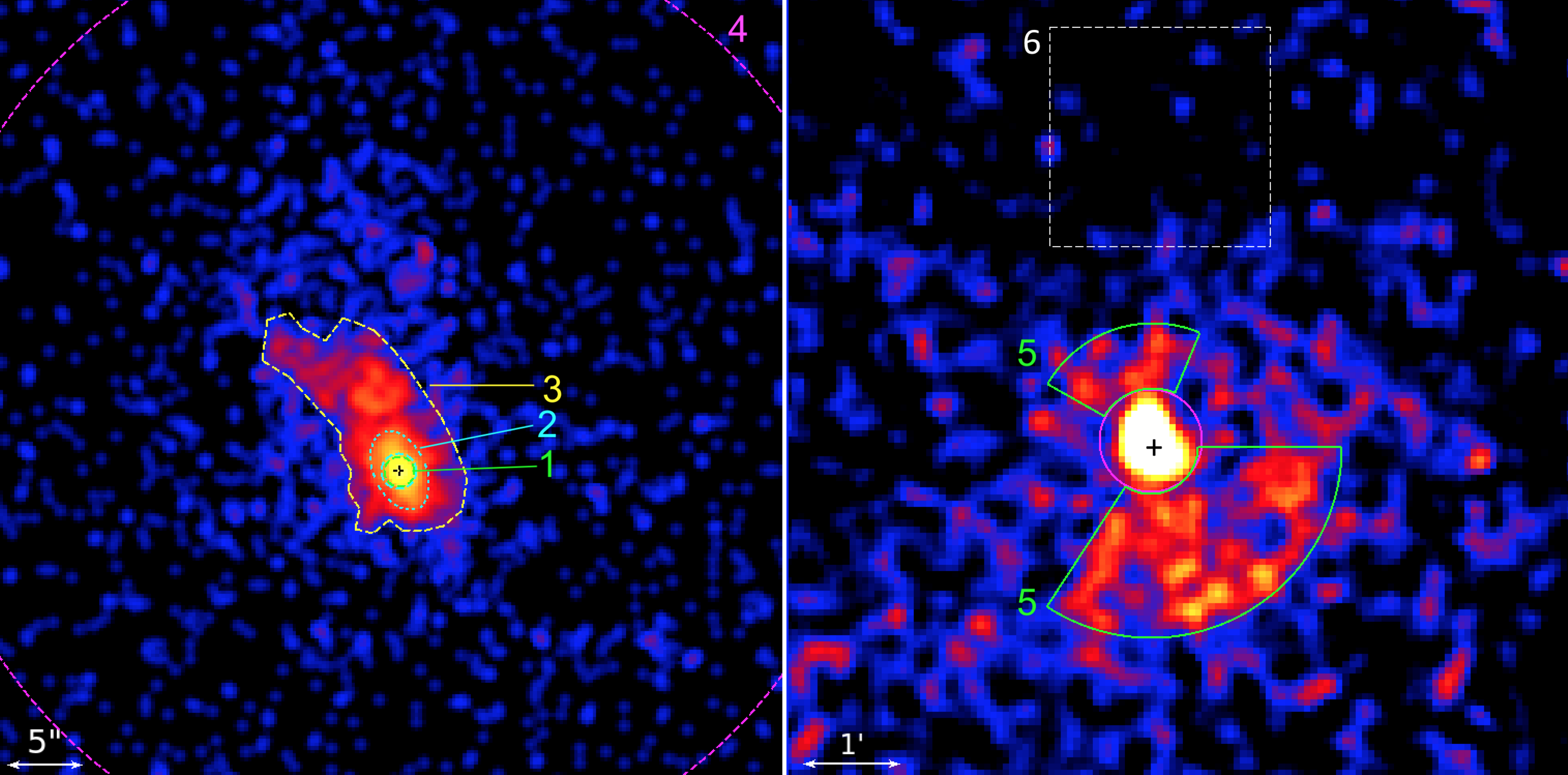}
\caption{Images obtained by merging data from all three ACIS observations (0.5--8 keV, 131 ks; best viewed in color).
{\sl Left:} Image of the CN, binned by a factor of 0.5 (pixel size $0\farcs25$), and smoothed with an $r=0\farcs75$ (3 pixel) Gaussian kernel.  
{\sl Right:}  Exposure-map-corrected image of the large-scale extended nebula, binned by a factor of 8 (pixel size $3\farcs94$) and smoothed with a 3 pixel ($r=11\farcs8$) Gaussian kernel.  
The black cross marks the pulsar position.  Shown are the following spectral extraction regions:
{\sl Left}:
1 -- pulsar (innermost green $1\farcs0$ circle), 
2 -- pulsar vicinity (area between the green circle and cyan ellipse), 
3 -- CN (area between the cyan ellipse and yellow polygon), and 
4 -- CN surroundings (area between the yellow polygon and magenta circle).
{\sl Right}:
5 -- extended nebula (green pie annulus);
6 -- region used for background subtraction.
}
\label{fig-images}
\end{figure*}

Figure \ref{fig-images} features merged images (from all 3 observations) at two different resolutions. 
The deeper image of the CN (left panel) shows that the brightest part of the CN in the immediate vicinity of the pulsar position is elongated along the same direction as the larger CN. 
The merged image of the three observations (Figure 2, right panel) allows us to detect the fainter large-scale extended emission farther from the pulsar than it was possible to in the 20 ks image analyzed by KP07. 
In contrast to the CN, the extended nebula (EN) is brighter and more extended southwest (SW) of the pulsar. 
The brighter part of the EN shows approximate symmetry with respect to the same axis as the CN.

\subsubsection{Spectra}
For spectral analysis we divide the merged PWN image into five regions shown in Figure \ref{fig-images}.  
To find the absorbing Hydrogen column density $N_{\rm H}=N_{\rm H,22}\times10^{22}$ cm$^{-2}$, we analyze the spectra of the bright inner part of the CN (region 3 in Figure \ref{fig-images}, which excludes the pulsar and its immediate vicinity ). 
We fit the unbinned spectra with an absorbed power-law (PL) using XSPEC's {\tt tbabs} absorption model (which uses absorption cross sections from Wilms et al.\ 2000), and obtained $N_{\rm H,22}=0.69\pm0.10$, $\Gamma=1.36\pm0.13$, and a W-statistic\footnote{The W-statistic is a generalization of the C-statistic (Cash 1979) developed for use in X-ray spectra with low counts and a background.  See https://heasarc.gsfc.nasa.gov/docs/xanadu/xspec/wstat.ps for details.} value of 1185.8 (for 1022 dof). 
To evaluate the fit, we used XSPEC's {\tt goodness} tool to perform 2500 Monte Carlo simulations of spectra using the best-fit parameters: in only 0.08\% of the realizations were the fits better by chance. 
Fitting the background-subtracted\footnote{The background accounts for only 2\% of the counts from this region.} binned spectra from the same region yields parameters comparable to, though less constrained than, the unbinned spectra.
The obtained $N_{\rm H,22}$ is consistent with $N_{\rm H,22}=0.72\pm0.14$ obtained by Kargaltsev \& Pavlov (2007), so we fix $N_{\rm H,22}=0.7$ for subsequent fitting.

The spectrum of the pulsar's immediate vicinity (region 2) is the hardest, with $\Gamma=1.20\pm0.11$.
The rest of the CN, its surroundings, and the large-scale extended emission all have somewhat softer spectra with very similar photon indices, $\Gamma\simeq1.5$ (see Table \ref{spatially-resolved-spectra} for details). 
In particular, there is no significant difference between the spectrum of the SW and NE parts of the EN (the two parts of region 5) when fitted separately.

The J1809 CN exhibits variability between observations 14820 and 16489, i.e., on timescales $<8$ months (see Figure \ref{fig-variability}).
However, the northern outflow spectrum does not vary significantly between the observations: $\Gamma_{3853}=1.35\pm0.18$, $\Gamma_{14820}=1.63\pm0.17$, and $\Gamma_{16489}=1.65\pm0.15$ (for the first, second and third observation, respectively).
The flux of region 3 is significantly lower in the latter two observations compared to the first one: $(1.72\pm0.13)\times10^{-13}$ versus $(1.14\pm0.08)\times10^{-13}$ and $(1.19\pm0.07)\times10^{-13}$ erg cm$^{-2}$ s$^{-1}$, respectively (see Table \ref{tab-variability} for details).

\begin{deluxetable*}{ccccccccccc}
\tablecolumns{4}
\tablecaption{Spectral fits for the PWN regions shown in Figure 2.
\label{spatially-resolved-spectra}}
\tablewidth{0pt}
\tablehead{\colhead{Region} & \colhead{Area} & \colhead{Counts} & \colhead{$\mathcal{B}$} & \colhead{$N_{\rm H,22}$} & \colhead{$\Gamma$} & \colhead{$\mathcal{N}_{-5}$} & \colhead{$\chi_\nu^2$ or $^\dagger$W-stat}  & \colhead{$F_{-13}$} & \colhead{$L_{31}$} 
}
\startdata
2 & 9.92 & $248\pm16$ & 1 & $0.87\pm0.23$ & $1.34\pm0.23$ & $0.82\pm0.25$ & $^\dagger203.4$ (265) & $0.66\pm0.06$  & $8.57\pm0.73$  \\ 
3 &  88.6  &  $815\pm29$  &  15  &  $0.86\pm0.21$  &  $1.64\pm0.18$  &  2.05--3.15   &  1.274  (43)  & 1.21--1.86 & 15.8--24.3  \\
4 & 3040 & $964\pm41$ & 15 & $0.48\pm0.15$ & $1.34\pm0.17$ & $1.72\pm0.39$ & 1.045 (90) & $1.37\pm0.09$ & $18.0\pm1.1$  \\
5 & 16935 & $1873\pm43$ & 50 & $0.64\pm0.16$ & $1.50\pm0.18$ & $4.6\pm1.1$ & 1.096 (87) & $3.16\pm0.22$ & $41.3\pm3.0$ \\
\hline
2  & 9.92 & $248\pm16$ & 1 & 0.7 & $1.20\pm0.11$ & $0.66\pm0.09$ & $^\dagger203.9$ (266) & $0.63\pm0.04$  &  $8.20\pm0.47$ \\ 
3 & 88.6 & $815\pm29$ & 15 & 0.7 & $1.53\pm0.08$ & 1.72--2.60 & 1.257 (44) & 1.14--1.72 & 15.5--22.5  \\
4  &  3040  &  $964\pm41$  &  15  &  0.7  &  $1.55\pm0.09$  &  $2.29\pm0.22$ & 1.055 (91) & $1.48\pm0.06$ & $19.3\pm0.8$  \\
5 & 16935 & $1873\pm43$ & 50 & 0.7 & $1.55\pm0.09$ & $4.99\pm0.44$ & 1.085 (88) & $3.23\pm0.13$ & $42.2\pm1.7$ 
\enddata
\tablenotetext{}{
Spectral fits for the extended emission.  
Listed are the region number, area (in arcsec$^2$), net counts, minimum counts bin$^{-1}$ $\mathcal{B}$, absorbing Hydrogen column density $N_{\rm H,22}$ in units of $10^{22}$ cm$^{-2}$, photon index $\Gamma$, and PL normalization $\mathcal{N}_{-5}$ in units of $10^{-5}$ photons s$^{-1}$ cm$^{-2}$ keV$^{-1}$ at 1 keV.
The spectra were fitted using $\chi^2$ statistics, unless noted by $^\dagger$ in which case W-statistics were used.
The unabsorbed flux, $F_{-13}$, and luminosity, $L_{31}$, are given in units of $10^{-13}$ erg cm$^{-2}$ s$^{-1}$ and $10^{31}$ erg s$^{-1}$, respectively, in 0.5--8 keV.  The luminosity is calculated for $d=3.3$ kpc.
In the first set of fits (above the horizontal line), $N_{\rm H,22}$ was treated as a free parameter, and in the second set (below the horizontal line), $N_{\rm H,22}$ was fixed at 0.7. 
Due to low counts, region 2 was fitted using W-statistics (noted with $^\dagger$), and $\chi^2$ statistics were used for all other regions. 
For region 3, the normalizations were untied in the different data sets (due to the emission variability), therefore the ranges of normalizations, fluxes, and luminosities are listed instead of the single values with their  1$\sigma$ uncertainties, as  for the other regions. 
}
\label{tbl-spectra}
\end{deluxetable*}

\begin{deluxetable*}{lccccccccc}
\tablecolumns{4}
\tablecaption{Spectral Fits for the Variable CN Regions
\label{tab-spectra-variable-region}}
\tablewidth{0pt}
\tablehead{\colhead{ObsID} & \colhead{Area} & \colhead{Counts} & \colhead{$N_{\rm H,22}$} & \colhead{$\Gamma$} & \colhead{$\mathcal{N}_{-5}$} & \colhead{W-stat (dof)}  & \colhead{$F_{-13}$} & \colhead{$L_{31}$} }
\startdata
3853  &  81.1  &  $157\pm13$  &  0.7  &  $1.40\pm0.16$  &  $1.85\pm0.29$  &  87.1 (114)  & $1.40\pm0.11$  & $18.2\pm1.5$  \\
14820  &  103.9  &  $180\pm13$  &  0.7  &  $1.51\pm0.15$  &  $1.34\pm0.21$  &  128.1 (131)  &  $0.90\pm0.06$   & $11.8\pm0.8$  \\
16489  &  56.64  &  $213\pm15$  &  0.7  &  $1.47\pm0.13$  &  $1.07\pm0.15$  &  129.1 (158)  &  $0.75\pm0.05$  &  $9.8\pm0.6$ 
\enddata
\tablenotetext{}{
Fits to the unbinned spectra of the variable part of the CN (regions shown in Figure \ref{fig-variability}). 
The notations are the same as in Table 3.}
\label{tab-variability}
\end{deluxetable*}

\subsection{Pulsar}

\subsubsection{Motion}
In an attempt to measure the pulsar's proper motion, we compared the positional changes of the pulsar position between the three observations.  

We used {\tt wavdetect} to determine the positions (and uncertainties) of 13 reference sources detected in all three epochs within $3\farcm5$ of the pulsar. 
These positions were used to determine mean shifts (weighted means of the shifts for individual sources) between the images and the corresponding uncertainties.
The shifts in both directions (R.A. and Decl.) are quite substantial, up to $1\farcs2$ (in Decl.) between the second and thirds observations. 
The uncertainties of the mean shifts are in the range of $64-69$ mas.
  
As a next step we measured the position of the pulsar in each observation.  
Since {\tt wavdetect} can be confused by the presence of the CN surrounding the pulsar, we manually calculated the centroid positions of all counts within $r=1\farcs25$ of the brightest pixel for each observation and the corresponding uncertainties (80-100 mas). 
We then fitted (by $\chi^2$ minimization) the steady motion model to the measured positions taking into account the calculated mean shifts between the images from different observations. 
The error budget included the uncertainty of the pulsar position in each image as well as the uncertainties of the mean shift (which, in turn, include the positional uncertainties of the individual reference sources).

We found that the pulsar moves north with proper motion $\mu_\delta=(39.6\pm8.8)$ mas yr$^{-1}$.
The motion in the east-west direction was not significant, $\mu_\alpha=(-0.2\pm8.3)$ mas yr$^{-1}$. 
Although $\mu_\delta$ is formally significant, we caution that our measurement does not include any systematic uncertainties which may dominate the statistical ones. 
The most obvious systematic uncertainty is associated with the pulsar position measurements in the presence of surrounding extended emission. 
In addition, there can be non-negligible uncertainties associated with the ACIS plate scale calibration and detector plane deformations that cannot be described by simple shifts. 
We expect to have a more reliable proper motion measurement when the currently ongoing series of monitoring observations of J1809 is completed. 

At $d=3.3$ kpc, the tentativly measured  proper motion corresponds to a projected velocity $v_\perp\sim600$ km s$^{-1}$ northward.
Although this velocity appears reasonable when considering the average 3D pulsar velocity ($\sim400\pm40$ km s$^{-1}$; Hobbs et al.\ 2005), such a velocity would imply highly supersonic motion if the pulsar is in the ISM, for which we see no evidence, such as a bow shock CN or an extended tail along the direction opposite  the proper motion direction. 
Alternatively, if the pulsar has not left its SNR, it would be subsonic (or at most, transonic), but no evidence of its parent SNR is seen either (see below). 
Future observations should provide more accurate measurements (or constraints) for the pulsar motion.

\begin{figure}
\epsscale{1.0}
\plotone{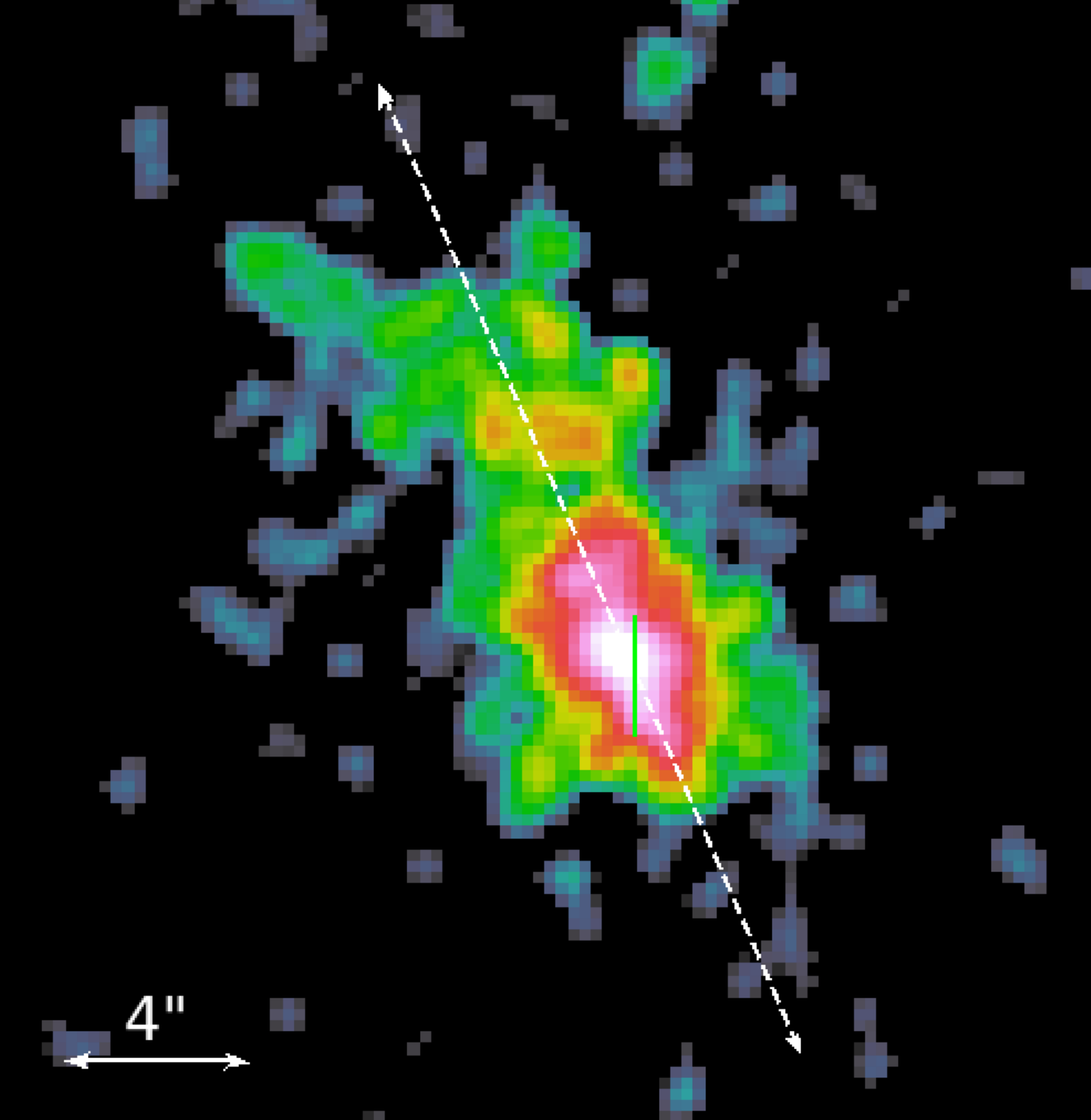}
\caption{Merged image of the CN showing the compact elongations interpreted as jets in Section 4.1.1.  The image was filtered to the 2.5--8 keV range (to remove most of the pulsar emission), binned by a factor of 0.5, and smoothed with a 3-pixel (r=$0\farcs74$) Gaussian kernel.  The white dashed line shows the apparent projected axis of the jets.  The narrow green ellipse (appearing as a line) marks the pulsar's radio position at MJD 51,506.
}
\label{fig-jets}
\end{figure}

\subsubsection{Spectrum}
To study the emission possibly originating from the pulsar\footnote{The emission from the brightest part of the PWN can, in principle, still be dominated by an unresolved bright PWN core rather than by the pulsar. A substantial pulsar contribution would be established either via detection of significant pulsations (which is not possible with these data) or by detecting the thermal component.}, we extracted the spectrum from an $r=1\farcs0$ aperture centered on the brightest pixel (consistent with the pulsar radio position at MJD 51,506: 18 09 43.147(7), --19 17 38.1(13); Morris et al.\ 2002). 
The encircled energy fraction of an $r=1\farcs0$ radius centered on an on-axis point source at 1.5 keV is $\approx$90\% (see Figure 4.6 of the {\sl Chandra} Proposers' Observatory Guide\footnote{http://cxc.harvard.edu/proposer/POG/html/chap4.html}).
At most, only 2\% of the pulsar flux could leak into the surrounding area (region 2).  
We estimated that this leaked pulsar flux amounts to a contamination of $<2\%$ of region 2's total flux; thus the effect of the small region sizes on the spectral analysis of the nebula is negligible.

We first fit the unbinned spectrum using a PL model. 
To account for contamination from the surrounding CN nonthermal emission, we also included a second PL component with the slope fixed at $\Gamma_{\rm CN}=1.2$, the best-fit value for region 2.
The fit yielded $\Gamma_{\rm PSR}=4.5\pm0.5$, a 0.5--8 keV unabsorbed pulsar flux of $7.5_{-1.7}^{+2.5}\times10^{-14}$ erg cm$^{-2}$ s$^{-1}$, an unabsorbed nebular contribution flux of $(2.0\pm0.3)\times10^{-14}$ erg cm$^{-2}$ s$^{-1}$, and a W-statistic of 808 (1533 dof). 
Performing Monte Carlo simulations in the same manner as described above resulted the fits being better by chance in 21\% of the 2500 realizations. 
Despite the good quality of the fit, the large (albeit uncertain) value of $\Gamma$ is very unusual (middle-aged pulsars rarely exhibit PL spectra with slopes $\Gamma>2$; see Kargaltsev \& Pavlov 2008) and indicative of a thermal spectrum. 
Since we do not expect such a steep spectrum from the innermost region of the PWN, we conclude that the extracted spectrum is dominated by the pulsar emission.

Fitting the pulsar spectrum with a blackbody (BB) model (with $\Gamma_{\rm CN}$ fixed at 1.2) yielded $kT=174\pm20$ eV, a normalization corresponding to an equivalent emitting sphere of radius $R_{\rm BB}=0.84_{-0.26}^{+0.41}$ km, and a nebular contribution flux of $(4.0\pm0.4)\times10^{-14}$ erg cm$^{-2}$ s$^{-1}$. 
The quality of the fit was similarly good (with a W-statistic of 817 for 1533 dof).

\begin{figure}
\epsscale{1.0}
\plotone{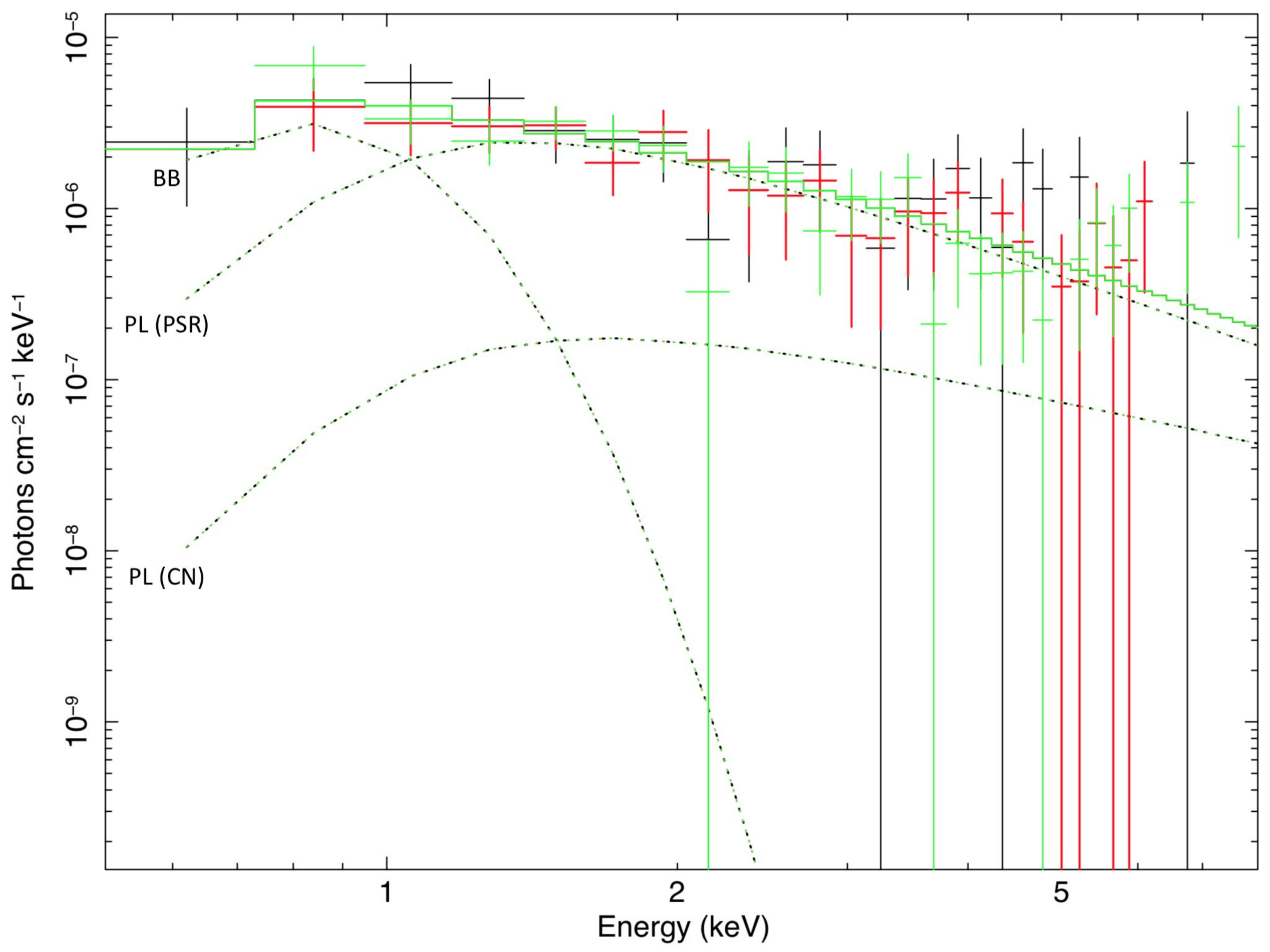}
\caption{Unfolded spectrum from the pulsar vicinity and the best-fit PL$_{\rm PSR}$+BB+PL$_{\rm CN}$ model with $\Gamma_{\rm PSR}$ fixed at 2.0, and $\Gamma_{\rm CN}$ fixed at 1.2. The spectrum is binned for visualization purposes only (see Section 3.3).}
\label{fig-pl-bb-spec}
\end{figure}

\begin{figure}
\epsscale{1.0}
\plotone{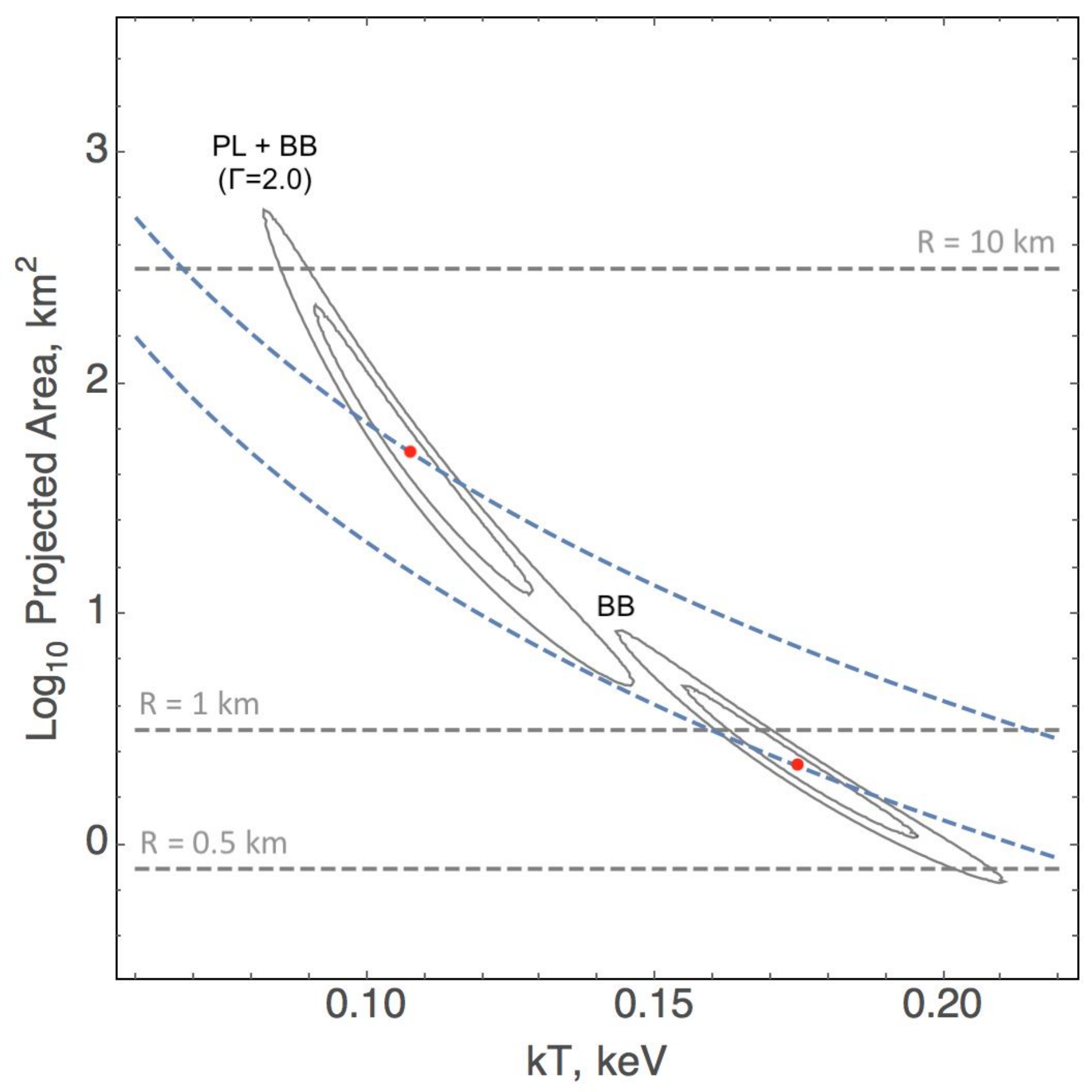}
\caption{Confidence contours (68\% and 90\%) for the PL+BB fit (top left contours, with $\Gamma_{\rm PSR}=2.0$) and BB fit (bottom right contours) to the pulsar spectrum. The horizontal dashed lines mark the projected areas corresponding to emitting BB equivalent sphere radii of 10, 1, and 0.5 km, and the dashed blue curves mark lines of constant bolometric luminosities of $2.8\times10^{32}$ and $8.4\times10^{31}$ erg s$^{-1}$, respectively.  $N_{\rm H,22}=0.7$ and $d=3.3$ kpc are assumed.}
\label{fig-contours}
\end{figure}

We also attempted to fit the pulsar spectrum with a PL+BB model (with $\Gamma_{\rm CN}$ fixed at 1.2), which yielded $\Gamma_{\rm PSR}=3.0\pm1.2$, $kT=95\pm28$ eV, a normalization corresponding to an equivalent emitting sphere with radius $R_{\rm BB}=5.7$ km (at $d=3.3$ kpc), and a W-statistic of 805 (1531 dof). 
The obtained $\Gamma=3.0$ seems unrealistically high, so we fixed $\Gamma_{\rm PSR}=2.0$ (since it is consistent with the best-fit $\Gamma$ within the uncertainties and representative of the slopes of middle-aged pulsars). 
The resulting fit yielded $kT=107^{+21}_{-16}$ eV, an equivalent emitting sphere radius $R_{\rm BB}=4.0_{-2.0}^{+4.2}$ km, and a nebular contribution flux of $5_{-5}^{+9}\times10^{-15}$ erg cm$^{-2}$ s$^{-1}$. 
The fit quality was good (with a W-statistic of 806 for 1531 dof).
In Figure \ref{fig-pl-bb-spec} we show the unfolded spectrum of the pulsar together with this best-fit BB+PL model.
Figure \ref{fig-contours} shows the confidence contours for the best-fit parameters of the thermal component in the BB and BB+PL fits.

\section{DISCUSSION}

The three observations of the J1809 PWN have revealed the unexpectedly strong variability. 
The improved counts statistics allowed us to resolve the compact structure of the PWN core and constrain its spectrum. 
Finally, the observations have provided sufficient exposure to resolve the faint large-scale emission in the merged image. 
We discuss each of these aspects in more detail below.

\subsection{PWN Morphology}

\subsubsection{Compact Nebula}
Because of the soft spectrum of the pulsar, the morphology of the CN in its immediate vicinity (i.e., region 2) can be better investigated by restricting an image to photon energies $>2.5$ keV, thus significantly reducing the contribution from the pulsar emission.
The 2.5--8 keV image, shown in Figure \ref{fig-jets}, reveals evidence of small-scale ($\approx$3$''$) extensions originating at the pulsar along a projected axis with a position angle $\approx 29^\circ$ East of North, which is approximately the same direction in which the CN emission appears to be elongated. 
There are two possible explanations for the observed elongations. 
It could be two jets (polar outflows) of comparable length and brightness (cf.\ the bright inner jets of the Vela PWN; Pavlov et al.\ 2001), or it could be a torus (equatorial outflow) viewed edge-on (cf.\ the PWN of PSR B0540--69, Gotthelf \& Wang 2000; see also Ng \& Romani 2008). 
The former interpretation would imply that the jets are weak and short enough to be barely distinguishable from the diffuse surrounding emission, while in the latter case the torus would be weak and/or very compact. 

If this structure is a torus, then it would be viewed edge-on or nearly edge-on, i.e., the viewing angle $\zeta$ between the line of sight and pulsar's spin axis would be $\sim 90^\circ$. 
With this kind of geometry, one would expect to see $\gamma$-ray pulsations (see Figure 16 in Romani \& Watters 2010) unless the magnetic inclination angle, $\alpha$, is small. 
However, $\gamma$-ray pulsations from J1809 have not been reported, while a small value of $\alpha$ should result in stronger jets relative to the torus (see Figure~4 of B{\"u}hler \& Giomi 2016; a possible example of such PWN is the one powered by PSR B1509--58). 
In addition, if it were a torus, then the elongated CN would be oriented in the pulsar's equatorial plane, and the larger northeast extent could only be explained if the pulsar were moving southwest, in contradiction to the likely direction of proper motion.  
Therefore, we consider this interpretation unlikely.

The alternative interpretation implies relatively bright jets compared to the torus. 
This is possible if $\alpha$ is relatively small. 
It would also explain the lack of $\gamma$-ray pulsations. 
However, since the pulsar is seen in radio (although it is rather faint), the viewing angle $\zeta$ cannot be very large. 
This should cause one of the jets (the approaching one) to be brighter than the other (receding) one, which is not observed.

Both elongations appear to have similar brightnesses in the vicinity of the pulsar. 
In the 0.5--8 keV range, the northeastern and southwestern halves of region 2 (which encloses the elongation) contain $191\pm14$ and $174\pm13$ counts, respectively, and in the 2.5--8 keV range, they contain $100\pm10$ and $88\pm9$ counts. 
The lack of significant Doppler boosting implies that we are viewing the pulsar from a line of sight not too far offset from equatorial plane.
To resolve this apparent inconsistency, one could speculate that the bulk flow velocities in the jets of PWNe powered by small-$\alpha$ pulsars are lower compared to the bulk velocities observed in PWNe powered by more orthogonal (or near-orthogonal) rotators (e.g., the Crab and Vela PWNe).

The substantial variability of the CN's northeastern elongation supports the jet scenario. 
Indeed, {\sl CXO} observations of the Crab, Vela and PSR B1509--58 PWNe show that, on larger scales, jets are the most variable PWN elements. 
One should not forget that these are 3D structures often showing helix-like morphologies and large-scale bending (cf.\ the outer jet of the Vela PWN; Pavlov et al.\ 2003). 
Because of this, the flow in different parts of the jet may be approaching the observer with very different velocities, and the Doppler boosting factors can vary across the extent of a large-scale jet\footnote{This, however, requires maintaining at least a mildly relativistic flow speed.}. 
As the jet bends (or the helix rotates; see Durant et al.\ 2013 for discussion of both possibilities), different parts of the jet become more or less Doppler boosted. 
Additionally, knots in the jet (which can account for a significant fraction of jet flux) can form, brighten, and fade over time. 
This may explain the large morphological variations in the NW part of the CN as well as flux from the same region varying by a factor of $\sim$2 between the first and third observations. 
The lack of similarly variable bright structures in the opposite direction suggests the two jets experience different conditions.  
Such asymmetry is also observed for the outer jets of the Vela PWN, but it is not yet well understood (Pavlov et al.\ 2003; Durant et al.\ 2013); it is possible that (in both PWNe) the counter-jet is too faint (perhaps due to Doppler boosting) to observe variations.
It may be associated with the motion of the pulsar through the surrounding medium resulting in different ram pressures for the forward and backward jets. 
However, the Vela pulsar's transverse velocity of $\simeq65$ km s$^{-1}$ (Caraveo et al.\ 2001) implies transonic or even subsonic motion, while our estimate of the J1809 velocity implies a much greater ram pressure and a larger Mach number. 
 
At the tentative transverse velocity of $\sim 600$ km s$^{-1}$, the pulsar motion has surprisingly little impact on the morphology of the J1809 PWN. 
Indeed, PWNe of supersonically-moving pulsars typically exhibit cometary morphologies often accompanied by extended (parsec-scale) tails (see Kargaltsev et al.\ 2017b). 
We see no indication of these in the J1809 PWN. 
The apparent effects of ram pressure could be reduced if most of the pulsar's wind is channeled into polar outflows (e.g., in the case of a nearly-aligned rotator; see simulations by B{\"u}hler \& Giomi 2016), substantially increasing the bow shock stand-off distance compared to the cases of isotropic or equatorial wind.  
The hypothesis that most of the wind is channeled into polar outflows is in agreement with the apparent lack of torus in the J1809 PWN (see above). 
The direction of motion aligning with (or at least close to) the spin/jet axis would allow the wind to push the stand-off distance much further ahead of the pulsar than one would expect from isotropic or equatorial wind, thus explaining the large $\sim1'$ extent of the northern section of the extended nebula (region 5). 
Alternatively, the distance to J1809 can be overestimated which would reduce the pulsar velocity. 
A combination of the two factors is also possible.  
The current uncertainties (especially, if the systematic ones are included) in the tentative measurement of the proper motion allow for alignment between the PWN symmetry axis and the direction of the proper motion. Any firm conclusions about possible associations will need to await more reliable proper motion measurements.

\subsubsection{Large-scale PWN}
The asymmetry between the small- and large-scale PWN is puzzling:  the variable northeast side of the CN is brighter and larger than the southwest side, but the southwest side of the EN is more prominent (see Fig.~2). 
The contrasting morphology could be attributed to the pulsar motion. 
The ram pressure can have a greater impact on the unresolved, more isotropic pulsar wind component pushing it further behind the pulsar (the wind producing the diffuse emission surrounding the CN; i.e., region 4). 
Alternatively, the asymmetry of the large-scale PWN morphology could be explained if the magnetic field south of the pulsar is compressed due to an interaction with a denser ISM or a reverse shock inside the parent SNR. 
However, there is no observational evidence of the SNR, and the advanced spin-down age of the pulsar suggests that the SNR has already dissipated.

\subsection{Parent SNR}
\begin{figure*}
\epsscale{1.15}
\plotone{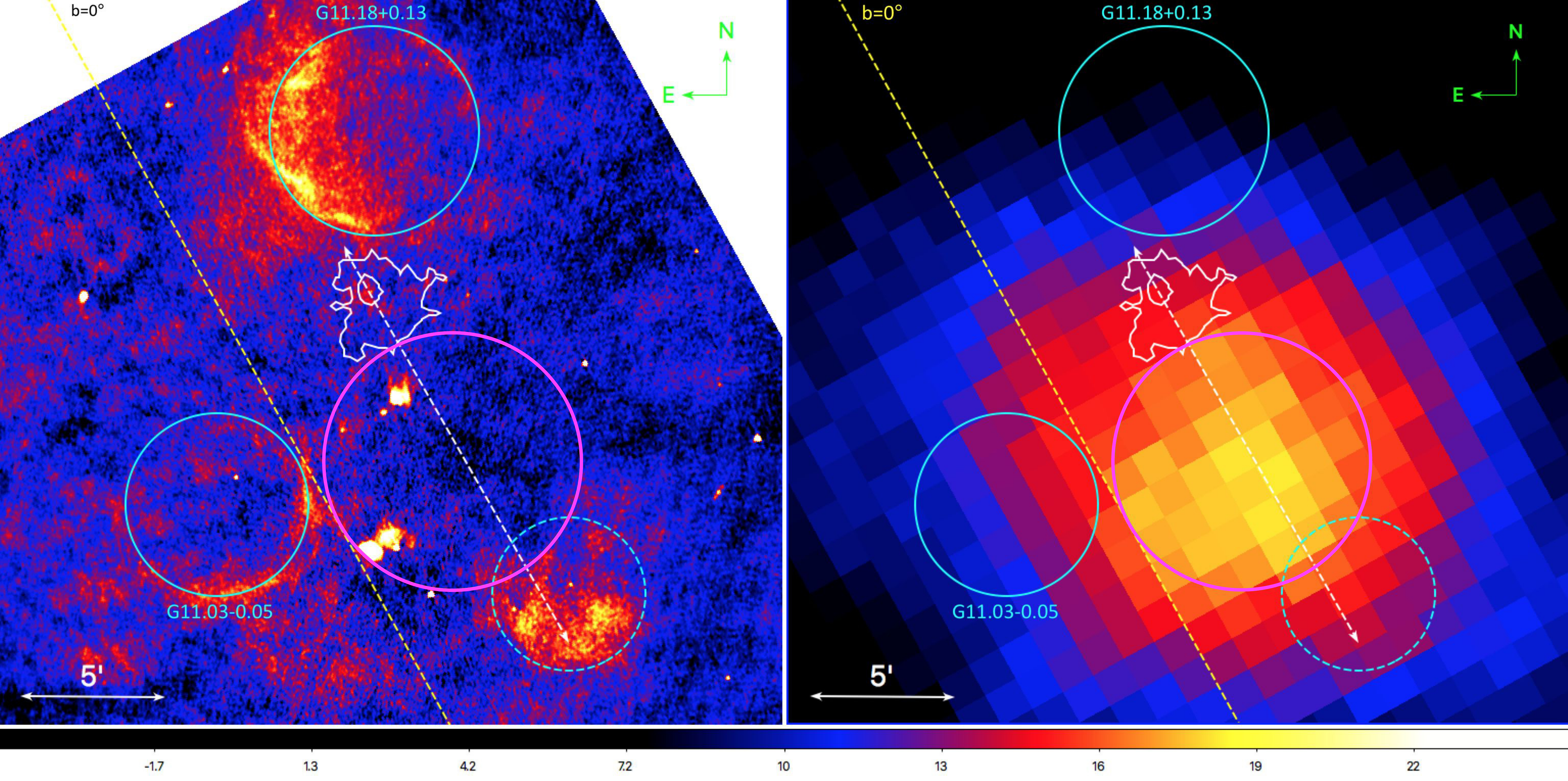}
\caption{
{\sl Left:} MAGPIS 1.5 GHz image of the J1809 field.  
{\sl Right:} TeV image J1809--193 from the HESS Galactic Plane Survey (H.E.S.S. Collaboration 2018); the color bar corresponds to the significance. 
In both images north is up and east is left, in equatorial coordinates; the dashed yellow line marks $b=0^\circ$ in Galactic coordinates.
Shown are the following: 
cyan circles (solid) -- the positions of SNRs G11.18+0.13 (northernmost) and G11.03--0.05 (easternmost); 
cyan circle (dashed) -- the position of candidate SNR G10.8750+0.0875 (westernmost; Helfand et al.\ 2006); 
white contours -- the X-ray extent of the J1809 PWN; 
magenta circle -- the peak TeV emission of HESS J1809--193; 
white dashed line -- the symmetry axis of the J1809 X-ray PWN.  
The sources of extended emission seen in the Eastern half of the green circle are HII regions (Castelletti et al.\ 2016).}
\label{fig-radio}
\end{figure*}

There are three incomplete radio shells in the  vicinity of J1809 (see Figure \ref{fig-radio}).
The shells north and southeast of J1809 (G11.18+0.13 and G11.03--0.05) have angular sizes of $\theta=6\farcm6$ and $\theta=7\farcm2$, respectively.
The angular size of an SNR in the Sedov phase can be approximated by $\theta_{\rm SNR} \approx 16'(E_{51}/n_0)^{1/5} t_4^{2/5} d_3^{-1}$ (Equation 1 in Romani et al.\ 2005), for a supernova energy $E_0=10^{51}E_{51}$ erg in an ISM of density $n_0$ cm$^{-3}$, at age $10^4 t_4$ yr and distance $3d_3$ kpc.
At the assumed distance of 3.3 kpc, $E_0=10^{51}$ erg, and $n_0=1$ cm$^{-3}$, the sizes of these shells would correspond to ages $<5$ kyr -- an order of magnitude smaller than the spin-down age of J1809.
Of these two, only G11.03$-$0.05 could be consistent with the tentative proper motion direction and its uncertainties.
If the pulsar was born in G11.03--0.05, it would have projected velocity $v_\perp=170\ (d/3.3\ {\rm kpc})(\tau/51\ {\rm kyr})^{-1}$ km s$^{-1}$ northwest (with both the north and west components $\approx 120$ km s$^{-1}$). 
Since such a velocity might be (barely) compatible with the tentatively measured $v_\perp$, given the possibility that the true pulsar age could be smaller, the association can not be firmly ruled out.

Another possible SNR, G10.8750+0.0875 (Helfand et al.\ 2006), lies about 13.5$'$  southwest of PSR J1809--1917.
The near-alignment between the PWN symmetry axis, TeV offset, and direction to the candidate SNR hint at an association, however, the angular size of the SNR (at an assumed distance of 3.3 kpc) is incompatible with the pulsar's age.
Even if J1809 is half as old as its spin-down age suggests, G10.8750+0.0875 is too small by a factor of a few. 
Finally, it is also possible that the parent SNR of J1809 is too cold and faint to be seen in the X-ray and radio images, which we tentatively consider as the most likely possibility.

\subsection{PWN and Pulsar Spectra}
The measured pulsar and PWN spectra are consistent with (albeit better constrained than) those reported by KP07.
At the DM distance of 3.3 kpc, the fluxes of the regions analyzed above yield a total PWN 0.5--8 keV luminosity $L_{\rm PWN}\sim1\times10^{33}$ erg s$^{-1}$, corresponding to a total PWN efficiency $\eta_{\rm PWN}\sim5\times10^{-4}$ -- a typical value for young Vela-like PWNe of similar ages (Kargaltsev \& Pavlov 2008).

The only appreciable spectral change seen across the extended emission is $\Delta\Gamma\approx0.35$ between the pulsar vicinity ($\Gamma=1.20\pm0.11$) and the CN ($\Gamma=1.55\pm0.09$).
No significant changes in the spectra are seen across the large-scale emission, up to $\sim2'$ (2 pc) from the pulsar.
The change of the CN spectrum in the pulsar vicinity can be understood if the magnetic field is high and rapid synchrotron cooling is occurring (e.g., in the Crab and Vela PWN tori the spectral slope changes by about the same amount). 
However, the lack of softening due to continued cooling in the extended nebula (up to parsec-scale distances) is surprising. 
This may indicate either low magnetic fields ($B\sim10$ ${\rm \mu}G$) and high bulk flow speeds, or particle re-acceleration far from the pulsar (e.g., due to magnetic reconnection). 
The lack of spectral softening across parsec-scale distances have been observed in PWNe before (e.g., it can be seen in tails produced by supersonically-moving pulsars; Klingler et al.\ 2016a,b).

\subsection{Association with HESS J1809--193}
The brightest part of HESS J1809--193 lies about 7$'$ southwest of PSR J1809--1917 (see Figure \ref{fig-radio}). 
Such large offsets between TeV and X-ray PWNe are typical, as the brightest parts of the TeV-emitting regions are powered by aged electron populations produced tens of kyr in the past (assuming typical ISM photon densities) when the pulsar's particle production rate was higher. 
Thus, these relic electron populations have had time to diffuse and advect further away from the pulsar, and/or were swept away when the PWN passed through a reverse shock in the past (see, e.g., Kolb et al.\ 2017).
The H.E.S.S. Collaboration (2017) reports that HESS J1809--193 fulfills all criteria for a TeV PWN association with PSR J1809--1917 (in considerations of pulsar-TeV source offset, TeV extension vs.\ age, TeV luminosity vs.\ $\dot{E}$, and surface brightness vs.\ $\dot{E}$). 
The potential association of HESS J1809 and PSR J1809 and its implications were discussed by KP07.
The improved positional accuracy of HESS J1809 has provided further support to their discussion by revealing that the X-ray/TeV offset vector aligns with the symmetry axis of the large-scale X-ray PWN morphology, and that the X-ray jets also align with this common axis, providing another link between the X-ray and TeV nebulae. 
For instance, the direction of the offset of the relic TeV PWN in the case of PSR B1706--44 is aligned with the pulsar's proper motion and with the symmetry axis of its large-scale X-ray PWN (HESS Collaboration 2011).
On the other hand, the direction of TeV PWN offset, the X-ray PWN symmetry axis, and the direction of pulsar motion need not to be aligned in a reverse shock scenario, even if the relic PWN (TeV source) and X-ray PWN are associated.
For instance, the relic PWN of the Vela pulsar lies far from the symmetry axis of the X-ray PWN because of a reverse shock (see Abramowski et al.\ 2012). 

It has been been proposed that diffuse unidentified GeV $\gamma$-ray emission in the vicinity of HESS J1809--193 is the counterpart of this TeV source, and that both the GeV and TeV sources originate from hadronic emission due to the interaction of the nearby SNRs G11.18+0.13, G11.03--0.05, G11.2--0.3, and G11.4--0.1 (the latter two are located outside the field of view of Figure \ref{fig-radio}) with molecular clouds in the region (Araya 2018).
However, the peak of the GeV emission is coincident with the position of the SNR/PWN G11.2--0.3 (roughly 20$'$ east from the TeV peak of HESS J1809--193), whose extent appears stretched in the directions of the other three SNRs. 
The HESS Galactic Plane Survey (GPS) image shows that HESS J1809--193 exhibits no hints of emission (e.g., local maxima or brightness enhancements) from any of the SNRs (see Figure \ref{fig-radio}), thereby making an association between HESS J1809--193 and the nearby SNRs unlikely.

\section{Summary}

We have presented an analysis of deep {\sl Chandra} observations of the PWN created by PSR J1809--1917. 
The images reveal a CN elongated along the Northeast-Southwest direction, extending further to the Northeast ($15''$) than it does to the Southwest ($5''$). 
The northeastern part of the CN shows variability, alternating the direction to which it curves on a timescale of a few months, making the J1809 PWN one of four PWNe known to exhibit variability (along with the Crab, Vela, and B1509--58 PWNe). 
The spectrum of the CN is well-described by an absorbed PL with $\Gamma=1.53\pm0.08$. 
We interpret the CN as a jet-dominated outflow viewed from a direction substantially offset from the equatorial plane. 
This interpretation is consistent with the lack of $\gamma$-ray pulsations, and also with the linear extensions originating from the pulsar position in the 2.5--8 keV image (likely a polar outflow), which exhibit a hard X-ray spectrum ($\Gamma=1.20\pm0.11$).

The pulsar's spectrum can be described by different models:
(1) a PL model (though with an unusually large $\Gamma\sim4$), (2) a BB model ($kT=174\pm20$ eV, $R\sim 0.8$ km, consistent with a hot spot on the neutron star surface), or (3) a PL+BB model ($\Gamma=2$, $kT=107_{-16}^{+21}$ eV, $R=4.0^{+4.2}_{-2.0}$ km at 3.3 kpc, consistent with 
nonthermal emission from the pulsar's magnetosphere plus thermal emission from the entire neutron star surface).
The data are not of high enough quality to confidently differentiate between the models.

The PWN also features fainter emission extending up to $\sim1'$ from the pulsar to the northeast and $\sim2'$ from the pulsar to the southwest (contrary to the CN which is longer to the northeast), along the same axis of symmetry as the CN (about $30^\circ$ east from north). 
This axis also aligns with the direction of offset of HESS J1809--193, which is likely the PWN's TeV counterpart. 
We see no evidence of spectral cooling across the extended emission, with the same X-ray spectral slope, $\Gamma\approx1.55\pm0.09$, virtually identical to that of the CN.  
Lacks of spectral changes up to parsec-scale distances have been observed in some pulsar tails created by supersonically-moving pulsars, but no morphological evidence of supersonic motion is seen in this PWN (e.g., no bow shock head or tail). 
This is at odds with the tentative measurement of a high projected pulsar velocity, $v_\perp \sim600$ km s$^{-1}$, directed northward.
The 0.5--8 keV luminosity of the entire PWN, $L_{\rm PWN}\approx 9\times10^{32}$ erg s$^{-1}$, corresponds to a PWN efficiency $\eta_{\rm PWN}\sim5\times10^{-4}$, typical of Vela-like pulsars. 

A series of deeper {\sl Chandra} observations will allow us to monitor and characterize the variability of the CN, discern the pulsar's spectrum, study the extent and spectrum of the faint extended emission, and better constrain the pulsar velocity.

{\em Facilities:} \facility{{\sl CXO}}, \facility{HESS}

\acknowledgements

We would like to thank Genavie Spence for her work on data analysis, and Samar Safi-Harb for the insightful discussions. 
We would also like to thank the referee for their helpful comments. 
Support for this work was provided by the National Aeronautics and Space Administration through {\sl Chandra} Award Number GO3-14072, issued by the {\sl Chandra X-ray Observatory} Center, which is operated by the Smithsonian Astrophysical Observatory for and behalf of the National Aeronautics Space Administration under contract NAS8-03060.

\end{document}